\documentclass[prl,twocolumn,superscriptaddress,floatfix,nopacs]{revtex4-1}
\usepackage{setspace}

\usepackage{amsmath,amsfonts,epsf,dcolumn,natbib,graphicx}
\usepackage{ascii}
\usepackage{comment}
\usepackage{xcolor}
\usepackage{lineno}
\newcommand{\be}{\begin{equation}}
\newcommand{\ee}{\end{equation}}
\newcommand{\beq}{\begin{equation}}
\newcommand{\eeq}{\end{equation}}
\newcommand{\bea}{\begin{eqnarray}}
\newcommand{\eea}{\end{eqnarray}}
\begin{document}
%\linenumbers
\bibliographystyle{plainnat}
%\draft

%\topmargin 0.1cm
%%%%%%%%%%%%%%%%%%%%%%%%%%%%%%%%%%%%%%%%%%%%%%%%%%%%%%%%
%%
\title{
  Elucidation of the subcritical character of the liquid--liquid transition in dense hydrogen
%using large-scale {\it ab-initio} simulations
}

\author{Valentin V.~Karasiev}
\email{vkarasev@lle.rochester.edu}
\affiliation{
Laboratory for Laser Energetics,
University of Rochester,
250 East River Road,
Rochester, New York 14623 USA
}
\author{Joshua Hinz}
\affiliation{
Laboratory for Laser Energetics,
University of Rochester,
250 East River Road,
Rochester, New York 14623 USA
}
\author{S. X.~Hu}
\affiliation{
Laboratory for Laser Energetics,
University of Rochester,
250 East River Road,
Rochester, New York 14623 USA
}
\author{S.B.~Trickey}
\affiliation{
  Quantum Theory Project,
  Dept. of Physics,
University of Florida,
Gainesville FL 32603 USA
}

\date{Dec. 26, 2020} %; vers.\ v8c; {\it NOT} for circulation outside LLE \& UF}

\maketitle

\renewcommand{\baselinestretch}{1.05}\rm

%\begin{spacing}{2.0}
%\setstretch{2.0}
%

\noindent {\bf ARISING FROM Bingqing Cheng et al. \emph{Nature}
%https://doi.org/10.1038/s4158610-11020-2677-y (2020)}
https://doi.org/10.1038/s41586-020-2677-y (2020)}

Determining the liquid--liquid phase transition (LLPT) in
high-pressure hydrogen is a longstanding challenge 
with notable variation in experimental and calculated
results. See Refs. 
%\onlinecite{DejarlaisKnudsonRedmer2020,
\onlinecite{GregoryanzEtAL2020,HinzEtAl2020,RilloEtAl2019,Lu..Dai.CPL.2019} 
and works cited therein for both calculational and experimental developments.  
Until recently, the 
computational consensus was for a first-order transition.  Calculated values 
differed but, for example, our results on $700  \le T \le 3000$ K are a
curve along $320 \ge P \ge 70$ GPa \cite{HinzEtAl2020}. 
Driven by molecular H$_2$ dissociation, transition signatures include
density jumps, qualitative and sharp changes in ionic pair
correlation functions (PCFs), and abrupt dc conductivity and reflectivity changes.   Coupled-electron ion Monte Carlo
(CEIMC) %\cite{Morales..Ceperley.PNAS.2010,Morales..Ceperley.PRL.2013,
\cite{Pierleoni..Ceperley.PNAS.2016} results concur at least roughly
with those from  {\it ab-initio} molecular dynamics 
driven by consistent density functional theory (MD-DFT)
\cite{HinzEtAl2020} and show reasonable agreement with
experiment also.

In marked contrast, Cheng et al. \cite{Cheng..Ceriotti.N.2020} found
a continuous transformation from a molecular to an atomic
liquid that goes supercritical above $P \approx 350$ GPa,
$T \approx 400$ K.  They used MD driven by a machine-learnt potential
(MLP). They attributed the dramatic differences versus MD-DFT to two
causes. One is 
finite size effects that foster the formation of defective solids,
with the common use of $NVT$ dynamics tending to increased defect
concentration compared to that from the $NPT$ ensemble.
The other  is much shorter simulation times in the MD-DFT and CEIMC calculations
than in the MD-MLP ones.   

Those diagnoses implicate other issues.  Almost all of the MLP
training was on small systems ($N \le 108$ atoms).  That raises
questions of large-system transferability. Conceptually there is the
question of whether a single MLP can represent two
distinct chemical regimes (molecular, atomic) correctly.  The
straightforward way to test both the two diagnoses (system size and
duration limits) and their implications is against much larger, longer
MD-DFT calculations.  We have done such calculations and find that
neither diagnosis is sustained.

We investigated with $NPT$ MD
simulations driven by DFT forces with PBE exchange-correlation \cite{PBE}.
(PBE was used in Ref. \cite{Cheng..Ceriotti.N.2020} to train the MLP.)
We used system sizes from 256 through 2048 atoms per cell.  Brillouin zone
sampling used the Baldereschi mean value point   
for the simple cubic crystal structure 
$\mathbf{k}=(\frac{1}{4},\frac{1}{4},\frac{1}{4})$ \cite{Baldereschi.1973}.
{\sc Vasp} \cite{vasp3,vasppaw} was used for 1024 and
2048 atom systems, while the i-PI interface \cite{i-PI2.0} with {\sc Quantum Espresso} \cite{QE.2017} was used for 
256 and 512 atoms.
Consistent results from the two shows that the MD code and
technical choices (thermostat, barostat, etc.) are inconsequential.

\begin{figure*}%[H]
\includegraphics*[width=15.6cm,angle =00]{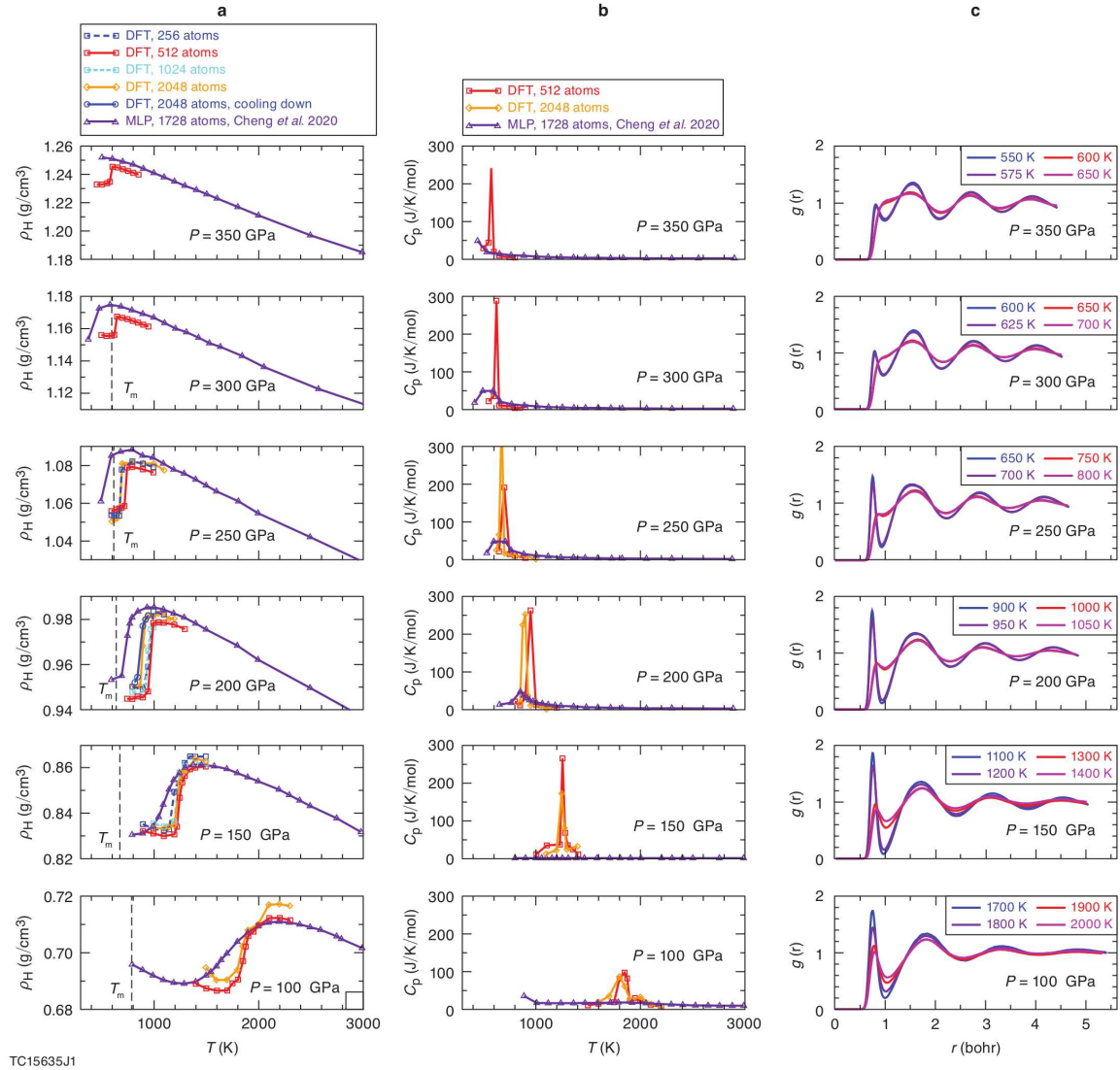}
\caption{
Comparison of MD results from the PBE exchange-correlation-based 
machine-learning potential (MLP) and {\it ab initio} MD-DFT (DFT) 
$NPT$ simulations.
\underline{Left column panels (a):} Hydrogen density as function of $T$ along six isobars.  
Melting temperature $T_{\mathrm{m}}$ for each isobar is shown by a vertical dashed line  \cite{Zha-Hemley2017}.
\underline{Middle column panels (b):} Molar heat capacity as a function of $T$ along the isobars.
\underline{Right column panels (c):} Pair correlation function (PCF) for each isobar
for two temperatures below the density jump and two above.  
}
\label{Fig1}
\end{figure*}

Our new large-system MD-DFT results agree with previous 
DFT-based and CEIMC simulations 
\cite{Lorenzen..Redmer.PRB.2010,HinzEtAl2020,RilloEtAl2019}: there is
a sharp molecular-to-atomic transition.  The qualitatively different
character compared with what comes from the MD-MLP is shown in
Fig. \ref{Fig1}. The left-column panels show density profiles
$\rho_{\mathrm H}(T)$ along isobars.  At 350 and 300 GPa, the
large-scale MD-DFT $\rho_{\mathrm{H}}(T)$ values jump $\approx$ 1\%
near $T= 650$ K. At 300 GPa, this is above the melting temperature
$T_{\mathrm{m}}$ \cite{Zha-Hemley2017}. In contrast, the 300-GPa
MD-MLP isobar has a steep density increase near $T= 500$ K  (in the
stable solid phase) \cite{Cheng..Ceriotti.N.2020}, but passes smoothly
through both that melt line and the LLPT.  Except for a systematic
offset, the MD-MLP $\rho_{\mathrm H}(T)$ matches the MD-DFT
$\rho_{\mathrm H}(T)$ in the atomic fluid region.

Figure \ref{Fig1} also shows unequivocally that there are \emph{no important
  finite-size effects} on the calculated LLPT. The density profiles on
each of the isobars  
($P=250$, $200$, $150$, and $100$ GPa) are almost identical
\emph{irrespective of atom count} (256, 512, 1024, or 2048).
The transition character is insensitive to system
size and specific technical choices of the MD code used, 
while the transition temperature $T_{\mathrm{LLPT}}$ is affected only
modestly. At $P=200$ GPa, for example, going from 256 to 2048 atoms
decreases $T_{\mathrm{LLPT}}$ by less than 100 K;
$\rho_{\mathrm{H}}$ values jump $\approx$ 3\% in MD-DFT simulations
for all system sizes.
A 512 atom system seems 
adequate to eliminate any  major finite-size effects.  This
outcome agrees with Ref. \cite{Geng..Ackland.PRB.2019}.  Those authors
found that four well-defined molecular shells in the PCF of a 3456-atom system 
were captured quite well in a 500-atom supercell calculation.

The molar heat capacity from MD-DFT as a function of $T$ is shown in
Fig. \ref{Fig1}, middle column. All the isobars 
exhibit divergent heat capacity character across the transition. 
They confirm that  
finite-size effects on $T_{\mathrm{LLPT}}$ are small and do not modify
that character. 

Figure \ref{Fig1} right-hand column shows the PCF on each isobar at
pairs of temperatures below and 
above the density jump.  Above, the first PCF peak virtually disappears,
confirmation of the density jump being in conjunction with the molecular
dissociation \cite{HinzEtAl2020}.

To test possible long simulation duration effects on $T_{\mathrm{LLPT}}$ or its
character, we did up to six sequential MD-DFT runs of roughly 1.8-ps 
duration each for a total of $\approx$10-ps duration.  This was at
200 GPa with 512 and 2048 atoms.  There were no meaningful differences
in the results in either case.  This outcome agrees with the results
of Geng et al. \cite{Geng..Ackland.PRB.2019} who did runs up to
6 ps and found no meaningful differences with respect to 1.5 ps (after
equilibration).

To investigate whether the nanosecond timescale might make the 
simulated transition smooth, we performed a set of
2048-atom MD-DFT $NPT$ simulations beginning with the atomic fluid
at 200 GPa.  Starting at 950 K, we cooled the system in 
sequential runs to 899, 849, and 824 K with
simulation durations around 8 ps for each temperature.
If the nanosecond timescale were to yield a smooth transition,
the hydrogen density during such a fast cooling curve would not 
drop sharply below the hypothetical smooth long-duration curve. But, as
evident in the Fig. \ref{Fig1} density plot at 200 GPa (left column),
the cooling curve (thin blue curve, circles), is almost identical to the one 
from MD simulations when the molecular fluid $T$ is
increased gradually (sharp transition shown by the solid
orange curve).

% Figure 2
\begin{figure}%[H]
\includegraphics*[width=8.4cm,angle =00]{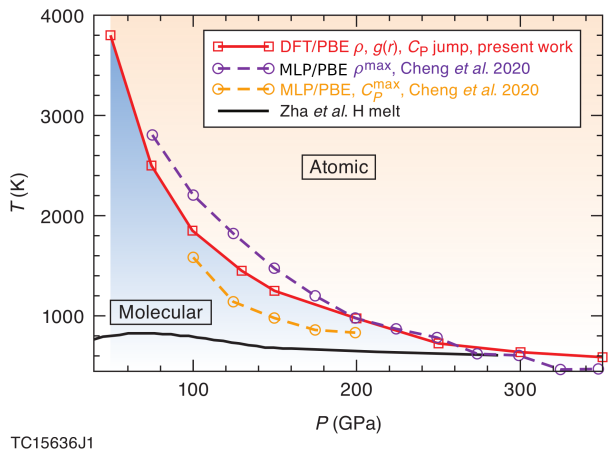}
\caption{
	The LLPT boundary from the present large-scale MD-DFT (DFT/PBE) simulations
	compared to MLP (MLP/PBE) $C_{P}^{\mathrm{max}}$ and $\rho^{\mathrm{max}}$ curves.
}
\label{Fig2}
\end{figure}

Figure \ref{Fig2} shows the LLPT curves associated with density jumps,
heat capacity peaks, and PCF peak disappearance.  For the new large-scale
MD-DFT calculations, those three criteria give one curve (virtually identical $P,T$
values), shown in red with squares at data points.  Two MD-MLP curves 
emerge from the analysis, however, one for the 
location of molar heat capacity maxima $C_{P}^{\mathrm{max}}$,
and another for the maximum density, $\rho^{\mathrm{max}}$. Consistent with
the foregoing discussion, there are striking differences.  
The MLP $C_{P}^{\mathrm{max}}$ curve lies well below the MD-DFT curve.
The MLP $\rho^{\mathrm{max}}$ curve is flatter than the MD-DFT reference
curve and lies close to it
%only for $P$ above about 170 GPa and then again at about
%P=70 GPa, T=2800 K.
only at about $P=70$ GPa, $T=2800$ K and then again for $P$ between about 170 and 300 GPa.

Given that neither the finite-size effect nor simulation duration
diagnosis advanced by Cheng et al. \cite{Cheng..Ceriotti.N.2020}
is sustained by our direct exploration, the remaining plausible cause
of the different physics they found must be in the MLP. The detailed origin of
that different physics is a bit obscure.  However, as discussed in our
Supplemental Information, documentation in the Supplemental
Information to Ref. \onlinecite{Cheng..Ceriotti.N.2020} confirms that
the MLP does not reproduce the behavior (be it physical or not) of
several MD-DFT calculations. Those differences, in addition to the stark LLPT
differences discussed here, confirm that the MLP is not systematically
related to the physics of a well-defined Born--Oppenheimer electronic
structure treatment of the H system.   The MD-MLP results 
instead are consistent, at least, with the MLP being a single interpolative,
approximate representation of the electronic structure of two
chemically distinct regimes (molecular, atomic) of the hydrogen
liquid.  

We conclude that the numerical evidence for supercritical behavior of
high-pressure liquid hydrogen based on the approximate MLP simulations
is unsupported by MD-DFT simulations on much larger systems for
significantly longer durations.  The diagnosis of the
difference between MD-MLP and MD-DFT calculations as being from 
size and duration effects is mistaken.  Rather, the supercritical behavior 
found in the MD-MLP calculations seems plausibly to be an artifact of a
disconnect of the MLP from underlying electronic structure differences
inherent in the chemistry of the LLPT.

\noindent{\bf Data availability}
\\
The data that support the findings shown in the figures are available from the 
corresponding author upon reasonable request.
\\

%\end{spacing}

\noindent {\bf Acknowledgements} 
\\

This report was prepared as an account of work sponsored by an agency of the U.S. Government. 
Neither the U.S. Government nor any agency thereof, nor any of their employees, makes any 
warranty, express or implied, or assumes any legal liability or responsibility for the accuracy, 
completeness, or usefulness of any information, apparatus, product, or process disclosed, or 
represents that its use would not infringe privately owned rights. Reference herein to any 
specific commercial product, process, or service by trade name, trademark, manufacturer, or 
otherwise does not necessarily constitute or imply its endorsement, recommendation, or favoring 
by the U.S. Government or any agency thereof. The views and opinions of authors expressed herein 
do not necessarily state or reflect those of the U.S. Government or any agency thereof.

V.V.K., J.H., and S.X.H. were supported by the Department of Energy 
National Nuclear Security Administration Award Number DE-NA0003856 and 
US National Science Foundation PHY Grant No. 1802964.  S.B.T. was supported by 
Department of Energy Grant DE-SC 0002139. 
This research used resources of the National Energy Research Scientific Computing Center, 
a DOE Office of Science User Facility supported by the Office of Science of 
the U.S. Department of Energy under Contract No. DE-AC02-05CH11231.
Part of the computations were performed on the Laboratory for Laser
Energetics HPC systems.
\\

\noindent{\bf Author contributions} 
V.V.K. conceived the project initially and designed the study. %and wrote the initial manuscript. 
V.V.K. and J.H. performed the MD-DFT simulations and post-processed the data. 
V.V.K. wrote the initial manuscript.
S.B.T. revised the conception and scope.  
%V.V.K. and J.H. performed the MD-DFT simulations and post-processed the data. 
V.V.K. and S.B.T rewrote the manuscript.
All authors discussed the results and revised the paper extensively.

\noindent{\bf Conflict of interests} The authors declare that they have no conflicts of interest.

\noindent{\bf Additional information}
\\
\noindent{\bf Supplementary information} is available for this paper at
https://doi.org/10.1038/xx  \\
\noindent{\bf Correspondence} and requests for materials should be addressed to V.V.K.

\end{document}